\documentclass[11pt]{article}
\usepackage{amsmath}
\usepackage{amssymb}

\begin{document}
\title{Dynamics of chiral primaries in $AdS_{3}\times S^{3} \times T^
{4}$}

\author{Aristomenis Donos and Antal Jevicki
\
\\ Physics Department\\
Brown University\\
Providence, Rhode Island 02912, USA \\
}

\maketitle

\begin{abstract}
We study in more detail the dynamics of chiral primaries of the D1/D5
system . From the CFT given by the $S_{n}$ orbifold a study of
correlators resulted in an interacting  (collective) theory of chiral
operators. In $AdS_{3}\times S^{3}$ SUGRA we concentrate on general
1/2  BPS configurations described
in terms of a fundamental string .We first establish a correspondence
with the linerized field fluctuations and then present the nonlinear analysis.
We  evaluate in detail the symplectic form of the general degrees
of freedom in Sugra
and confirm the appearance of chiral bosons. We then discuss the
apearance of interactions and the cubic vertex,in correspondence with
the $S_{N}$ collective field theory representation.
\end{abstract}

\newpage

\section{Introduction}

Recently a  dynamical descriptio of $\frac{1}{2}$ BPS configurations
of $AdS_{5}\times S^{5}$ SUGRA in terms of a fermionic (droplet)
picture was acomplished\cite{CJR, B,LLM}. The fermion dynamics is
obtained from a reduction of $N=4$ Yang-Mills theory on $R\times
S^{3}$ in a manner similar to the holomorphic projection of the Hall
effect. Its bosonic (collective droplet) representation is recovered
completely by the ' bubbling' ansatz of Lin,Lunin and Maldacena
\cite {LLM} . The complete equivalence between two pictures is
established through evaluation of the flux and the energy
\cite{LLM}, \cite{Horava:2005pv, RP,Balasubramanian:2005kk, DJR,
Yoneya:2005si, Gauntlett:2005ww, deMello,Mandal:2005wv,
Okuyama:2005st} and also the symplectic form of the boundary degrees
of freedom\cite{MR}. This dynamical map represents a most direct
realization of the AdS/ CFT correspondence. It is expected to play a
central role in further (nonperturbative) studies and provide a
basis for further extensions \cite{DJR}.
  The dynamics of $\frac{1}{2}$ BPS
states of Supergravity on $AdS_{3}\times S^{3}$ is similarly of
substantial interest. The system represents $D1-D5$ branes and is of
relevance for microscopic study of black holes and the
implementation of the $AdS_{3}/CFT_{2}$ correspondence. Early
studies of the dynamics of chiral primaries were given in
\cite{JR,JMR} and in a series of papers \cite{Lunin:2000yv,
Lunin:2002fw}. In \cite{JMR} based on a study of three-point
correlators on the $S_{n}$ orbifold a cubic interaction hamiltonian
was developed. In comparison with the single chiral collective boson
of $AdS_{5}\times S^{5}$ here the set of chiral primary fields is
extended by a further(fermionic) structure associated with forms on
$T^{4}$ or $K^{3}$ representing the compactification manifold. These
fields summarize the collective effects of a simple 1d system which
will be summarized and reviewed in sect.II. Recent studies on the
Supergravity side have given further understanding of most general
1/2BPS configurations in $AdS_{3}\times S^{3}$ gravity side one has
the 'bubbling' ansatz describing $\frac{1}{2}$ BPS configurations.
This ansatz is essentially the long string solution of
\cite{Lunin:2001fv, Lunin:2001jy, Lunin:2002bj}.
\par
In the present work we will discuss further the hamiltonian
realization of the
correspondence for the $D1-D5$ system.We will study in some detail
the dynamics
of the $AdS_{3}\times S^{3}$ 'bubbling' ansatz in Supergravity and
perform a comparison
with the collective theory of chiral bosons.For this we first present
an evaluation
of the symplectic form for the most general BPS configuration and
establish the chiral boson
description of these degrees of freedom from supergravity. In this we
follow the
method established by Maoz and Rychkov. We then to study the
perturbative
expansion of the hamiltonian and demonstrate  at the quadratic and cubic
level a correspondence with the collective boson description of
\cite{JMR}.

\section{Collective field theory of chiral primaries}

We begin by giving a brief summary
of the SCFT on symmetric  product $S^N(X)$ and of the cubic(collective)
representation for chiral primaries introduced in [16].the
compactifiction
manifold $X$ is either $T^{4}$ or $K3$ . The SCFT  has
(4,4) superconformal symmetry in both cases. We will work with
$T^{4}$ for simplicity and most  of the results extend  simply to  $K3$.
\par
The field content of the theory consists of: $4N$ real free bosons $X^{a
\dot{a}}_{I}$ representing the coordinates of the torus , their
superpartners are $4N$ free fermions $\Psi^{\alpha\dot{a}}_{I}$, where
$I=1,..,N$, $\alpha,\dot{\alpha} =\pm$ are the spinorial $S^{3}$
indices, and $a,\dot{a}=1,2$ are the spinorial indices on $T^{4}$.
  Using the relation between the Fermi fields:
$\Psi^{\alpha\dot{a}\,\dagger} = \Psi_{\alpha\dot{a}}=
\epsilon_{\alpha\beta}\epsilon_{\dot{a}\dot{b}}
\Psi^{\beta\dot{b}}$, the field content of the theory is determined
to be $4N$ real free bosons and  $2N$ Dirac free fermions, producing a
total central charge $c=6N$.
\par
In the free  CFT on $T^4$ one has non-trivial $S_N$ invariant chiral
primary operators
which are  constructed in correspondence  with conjugacy
  classes of $S_N$ .
One has the basic twist operators for $n$ free bosons: $X_{I}, I=1..n$,
defined through the OPE :
\begin{equation}\label{opg}
\partial X_{I}(z)\sigma_{(1..n)}(0)=z^{{\frac{1}{n}}-1} e^{-{\frac{2
\pi i }
{n}}I}\tau_{(1..n)}(0)+..
\end{equation}
They impose the boundary
conditions :
\begin{equation}\label{bcond}
  X_{I}(z\,e^{2 \pi i}, \bar{z}\,e^{-2 \pi
i})=X_{I+1}(z,\bar{z}), \; I=1..n-1, \; X_{n}(z\,e^{2 \pi i},
\bar{z}\,e^{-2 \pi i})=X_{1}(z,\bar{z})
\end{equation}

These twist operators play a distinguished role in the chiral ring,
they can be used to generate the rest by using the ring structure.
In the correspondence with gravity in $ADS_3 \times S^3$ they are in
one-one correspondence with single particle states.
\par
Let us begin with the operators appearing in the
untwisted sector,
  $n=1$:
\begin{eqnarray}\label{untwp}
   & \omega^{(0,0)}=1, \\
   & \omega^{a}=\psi_{0}^{+ a}, \\
   & \omega^{\bar a}=\bar \psi_{0}^{+ a}, \\
   & \omega^{a,\bar b }=\psi_{0}^{+ a} \bar \psi_{0}^{+ b}, \\
   & \omega^{a a' ,\bar b }=\psi_{0}^{+ a} \psi_{0}^{+ a'}
     \bar \psi_{0}^{+ b}, \\
   & \omega^{a ,\bar b \bar b' }=\psi_{0}^{+ a} \bar \psi_{0}^{+ b}
     \bar \psi_{0}^{+ b'}, \\
   & \omega^{(2,2)}=\psi_{0}^{+ 1} \psi_{0}^{+ 2} \bar \psi_{0}^{+ 1}
     \bar \psi_{0}^{+ 2}.
\end{eqnarray}

  The complet set of  vertex
operators can be  written in terms of the $6$ free scalar fields and the
twist operators.One considers the cycle $(1..n)$ and defines
  $S_{n}$ invariant $6$ dimensional vector observables (left
and right) :
\begin{eqnarray}\label{vect}
& Y_{L}(z)={\frac{1}{n}}\sum_{I=1..n}~(X_{I~L}^{1}, X_{I~L}^{2},
X_{I~L}^{3},X_{I~L}^{4},
\phi_{I}^{1},\phi_{I}^{2})(z),\\
  & Y_{R}(\bar z)={\frac{1}{n}}\sum_{I=1..n}~(X_{I~R}^{1}, X_{I~R}^{2},
  X_{I~R}^{3},X_{I~R}^{4},\bar \phi_{I}^{1},\bar \phi_{I}^{2})(\bar
  z),
\end{eqnarray}
The simplest vertex operator consists of the twist
operator for all 6 fields and having momenta along the 2 extra
$\Phi$ dimensions:

\begin{equation}\label{vertg}
O_{(1..n)}^{(0,0)}(z,\bar{z})=e^{i
(k_{L}Y_{L}(z)+k_{R}Y_{R}(\bar z) ) }
              \sigma_{(1..n)}(\Phi,X)(z,\bar{z})
\end{equation}
where $k_{L}=(0,0,0,0,{\frac{n-1}{2}},{\frac{n-1}{2}})$,
$k_{R}=(0,0,0,0,{\frac{n-1}{2}},{\frac{n-1}{2}})$ represents the left
and right momenta in $6$ dimensions.  All the other chiral
vertex operators are obtained by combining the states appearing in
the untwisted sector
with the above twisted operators. One has  the following
construction:first introduce the
product
\begin{equation}\label{basc}
O_{(1..n)}^{A}(z,\bar z) \leftarrow
O_{(1)}^{A}(z,\bar z)~O_{(1..n)}^{(0,0)}(z,\bar z)
\end{equation}
where $A$ index takes care of the spinorial indices which already
fully appear at untwisted level.  It is useful to define a
basis of forms in the target space
$X$ and spanning its $H^{(1,1)}(X)$; we will denote them as
  $\omega^{r}_{a\bar{a}}$ where r counts the forms
(for example, $r=1..4$ for $T^{4}$, and $r=1..20$ for $K3$)) and
$a,\bar{a}=1,2$ and they are $X$ indices. Using this forms and
summing over all permutations, it is possible to describe the scalar
chiral primaries up to a normalization constant :
\begin{eqnarray}\label{twist}
&O_{n}^{(0,0)}(z,\bar{z})={\frac{1}{(N!(N-n)!n)^{\frac{1}
{2}}}}\sum_{h\in S_{N}} O_{h(1..n)h^{-1}}(z,\bar{z}),\\
   & O_{n}^{r}(z,\bar{z})={\frac{1 }{(N!(N-n)!n)^{\frac{1}{2}}}}
\sum_{h\in S_{N}}O_{h(1..n)h^{-1}}^{a,\bar{a}}
\omega^{r}_{a\bar{a}}(z,\bar{z}),\\
   & O_{n}^{(2,2)}(z,\bar{z})={\frac{1}{4}}\,{\frac{1}
{(N!(N-n)!n)^{\frac{1}{2}}}}\sum_{h\in S_{N}}
O_{h(1..n)h^{-1}}^{ab,\bar{a}\bar{b}}\epsilon_{ab}
\epsilon_{\bar{a}\bar{b}}(z,\bar{z}),
\end{eqnarray}
\par
Of special use in construction of the hamiltonian are normal modes
defined from the residues of the
above operators :
$a_{n}^{0},a_{n}^{r},a_{n}^{(2,2)}$ ,these obey
canonical commutation relations
\begin{eqnarray}
\left[a_{n},a_{m}^{\dag} \right]&=&\delta_{n,m}\\
\left[a^{r}_{n},a_{m}^{s\dag} \right]&=&\delta_{n,m}\delta_{r,s}
\end{eqnarray}
Their correlation functions were determined from the operator
products in [16] ,they
are given by
\begin{eqnarray}
\langle a^{0\dag}_{n+k-1} a^{0}_{k} a^{0}_{n}\rangle
&=&\frac{1}{2}\frac{1}{\sqrt{N}} \left(\left(n+k-1\right) n k
\right)^{\frac{1}{2}}\\
\langle a^{r\dag}_{n+k-1} a^{s}_{k} a^{0}_{n}\rangle
&=&\frac{1}{2}\frac{1}{\sqrt{N}} \left(\left(n+k-1\right) n k
\right)^{\frac{1}{2}}\\
\langle a^{\left(2,2\right)\dag}_{n+k-1} a^{\left(2,2\right)}_{k}
a^{0}_{n}\rangle &=&\frac{1}{2}\frac{1}{\sqrt{N}}
\left(\left(n+k-1\right) n k \right)^{\frac{1}{2}}\\
\langle a^{\left(2,2\right)\dag}_{n+k-1} a^{r}_{k} a^{s}_{n}\rangle
&=&-\frac{1}{2}\frac{1}{\sqrt{N}} \left(\left(n+k-1\right) n k
\right)^{\frac{1}{2}}\\
\langle a^{\left(2,2\right)\dag}_{n+k-1} a^{0}_{k} a^{0}_{n}\rangle
&=&\frac{1}{2}\frac{1}{\sqrt{N}} \left(\left(n+k-3\right) n k
\right)^{\frac{1}{2}}.
\end{eqnarray}
\par
From the above correlation functions we read off the collective
field Hamiltonian
\begin{eqnarray*}
H= \sum_{n>0} \left[na_{n}^{\dag}a_{n}+ na_{n}^{r\dag}a_{n}^{r} +
na_{n}^{\left(2,2\right)\dag}a_{n}^{\left(2,2\right)} \right]\\
+\frac{1}{\sqrt{N}}\sum_{n,k>0} \sqrt{nk\left(n+k-1\right)} \left[
a^{\dag}_{n+k-1} a_{k}a_{n} +  a^{\dag}_{k} a^{\dag}_{n}a_{n+k-1}
\right]\\
+\frac{1}{\sqrt{N}}\sum_{n,k>0} \sqrt{nk\left(n+k-1\right)} \left[
a^{r\dag}_{n+k-1} a^{r}_{k}a_{n} +  a^{r\dag}_{k}
a^{\dag}_{n} a^{r}_{n+k-1}\right]\\
+\frac{1}{\sqrt{N}}\sum_{n,k>0} \sqrt{nk\left(n+k-1\right)} \left[
a^{\left(2,2\right)\dag}_{n+k-1} a^{\left(2,2\right)}_{k}a_{n} +
  a^{\left(2,2\right)\dag}_{k}
a^{\dag}_{n} a^{\left(2,2\right)}_{n+k-1}\right]\\
+\frac{1}{\sqrt{N}}\sum_{n,k>0} \sqrt{nk\left(n+k-1\right)} \left[
a^{\left(2,2\right)\dag}_{n+k-1} a^{r}_{k}a^{r}_{n} +
  a^{r\dag}_{k} a^{r\dag}_{n}a^{\left(2,2\right)}_{n+k-1} \right]\\
+\frac{1}{\sqrt{N}}\sum_{n,k>0} \sqrt{nk\left(n+k-3\right)} \left[
a^{\left(2,2\right)\dag}_{n+k-3} a^{\dag}_{k}a_{n} + a^{\dag}_{n}
a^{\left(2,2\right)}_{n+k-3} a_{k}  \right]
\end{eqnarray*}
The Hamiltonian takes the form
\begin{equation}
H_{coll}=\frac{1}{6} \sum_{n_{1},
n_{2},n_{3}}\delta_{n_{1}+n_{2}+n_{3},0} \alpha_{n_{1}}
\alpha_{n_{2}} \alpha_{n_{3}}  + \sum_{n} \alpha_{n}
T_{-n}\left(S\right) + \cdots.
\end{equation}
\par
We see that the (0,0) chiral primary field  which is described by the
modes $a_{n}^{\left(0,0\right)}$ is self interacting with a cubic
interaction of a familiar collective fermion. The other fields which
we denote as $s_{n}^{R},\, R=1\ldots 5$ (related to the operators $a^
{r},\, a^{2,2}$ interact with $\alpha$. This interacting theory is
seen to be associated with the collective degrees of freedom
consisting of $N$ eigenvalue coordinates $\lambda_{i},\, i=1\ldots N$
and the fermionic (zero modes) $\psi_{i}^{a,\alpha}$ with $S_{N}$
symmetry. The collective fields given by
\begin{equation*}
\Phi^{\left(p,q\right)}\left(x\right)=\sum_{i=1}^{N}\delta\left(x-
\lambda_{i}\right)W^{\left(p,q\right)}\left(\psi_{i},\bar{\psi}_{i}
\right)
\end{equation*}
where $W^{\left(p,q\right)}$ are associated with the $\left(p,q\right)
$ forms on $T^{4}$ or $K_{3}$. In particular the form $\left(0,0
\right)$ , $W^{\left(0,0\right)}=1$ leads to
\begin{equation*}
\Phi\left(x\right)=\sum_{i=1}^{N}\delta\left(x-\lambda_{i}\right)
\end{equation*}
which is just the density of eigenvalues, we denote the other
collective fields as $S^{R}$ \footnote{At the orbifold point the $O
\left(4\right)$ symmetry is manifestly visible while the $O\left(5
\right)$ must have a nonlinear origin}.
\par
The structure of collective fields is then recognized as the one of
the matrix-vector model. We have in general the density fields and
the current which separate in terms of chiral components
\begin{eqnarray*}
\Phi\left(x\right)&=& y_{+}\left(x\right)-y_{-}\left(x\right)\\
J^{\alpha}\left(x\right)&=& J^{\alpha}_{+}\left(x\right)- J^{\alpha}_
{+}\left(x\right)
\end{eqnarray*}
with the Hamiltonian
\begin{eqnarray*}
H= \int dx \left[\frac{1}{3}\left(y^{3}_{+}- y^{3}_{-}\right)+x^{2}
\left(y^{3}_{+}- y^{3}_{-}\right) +y_{+}J_{+}^{2}+y_{-}J_{-}^{2}+L^
{\prime} \right]
\end{eqnarray*}
where $L^{\prime}$ stands for higher order (quartic) coupling
. The commutation relations are those of a $U
\left(1\right)\times U\left(1\right)$ chiral boson
\begin{equation*}
\left[y_{\pm}\left(x\right),y_{\pm}\left(x^{\prime}\right) \right]=
\pm 2\delta^{\prime}\left(x-x^{\prime}\right)
\end{equation*}
and a Kac-Moody algebra.It represents a collective
field theory of a matrix-vector type studied in \cite{AJL}.
\par
It is simple to see how this general Hamiltonian encompases the
interaction of
chiral primaries written down above . First, the presence of the $x^
{2}$ potential induces a classical background
\begin{equation*}
\phi_{0}\left(x\right)=\frac{1}{\pi}\sqrt{\mu -x^{2}}
\end{equation*}
which is interpreted as a one-point function in the CFT collective
field theory (in later section we will reproduce this structure in
SUGRA). After the shift
\begin{equation*}
\phi\left(x\right)=\phi_{0}+\sigma\left(x\right)
\end{equation*}
one obtains the cubic interaction
\begin{equation*}
H_{3}=\int d\tau \frac{1}{\phi_{0}^{2}}\sigma\left(x \right)^{3}
\end{equation*}
where $\tau$ is the time of flight coordinate. In terms of modes
\begin{equation*}
H_{3}=\sum_{n_{1},n_{2},n_{3}}f\left(n_{1}+n_{2}+n_{3}\right)\alpha
\left(n_{1}\right)\alpha\left(n_{2}\right)\alpha\left(n_{3}\right)
\end{equation*}
with the form factor
\begin{equation*}
f\left(n\right)=\frac{n}{\sinh \pi n}.
\end{equation*}
For the special case of observables (corresponding to chiral
primaries) where one has momentum conservation $n=n_{1}+n_{2}+n_{3}=0
$ we see
\begin{equation*}
f\left(n\right)\rightarrow 1
\end{equation*}
resulting in a conserving, form factor . We emphasize that the
collective field theory (and also gravity) have a more general set of
generators.
What we have found is a  theory of a fermionic droplet interacting with
current degrees of freedom living on the boundary of the droplet.
\section{Linearized fields from SUGRA on $AdS_{3}\times S^{3}$}

In this section we will begin our discussion of the supergravity
side by a usefull linearized analysis. At the linearized level the
chiral
primaries have been identified in \cite{Sezgin}. The
effective action of the degrees of freedom, that we are interested
in, was given at the cubic level in \cite{Mihail, Arutyunov:2000by}, the pp limit was studied in \cite{Cremonini:2004kh}. We will identify  these chiral primaries by expanding the exact nonlinear solution of Sugra constructed in \cite{LMM}. 
\par
We consider the action of IIB supergravity in the string frame
\begin{equation*}
S=\int dx^{10}
\sqrt{-g}\left[e^{-2\Phi}\left(R+4\left(\nabla\Phi\right)^{2}\right)-
\frac{1}{12}H^{2} \right]
\end{equation*}
where $H$ denotes the field strength of the RR two form $H=dC^{\left(2
\right)}$. The
D1-D5 system is described by the the profile of a curve
$F_{i}\left(u\right),\,u=0\dots 2\pi Q_{5},\,i=1,\ldots,4$.

The solution in the 10D string frame is given by
\begin{eqnarray*}
&ds^{2}=\frac{1}{\sqrt{f_{1}f_{5}}}\left[-\left(dt- A_{i}dx^{i}
\right)^{2}
+ \left(dy +B_{i}dx^{i} \right)^{2} \right]+ \sqrt{f_{1}f_{5}}
d\mathbf{x}^{2}+ \sqrt{\frac{f_{1}}{f_{5}}} d\mathbf{z}^{2}\\
&e^{2\Phi}=\frac{f_{1}}{f_{5}}\\
&C^{\left(2\right)}_{ti}=\frac{B_{i}}{f_{1}},\,\,
C^{\left(2\right)}_{ty}=\frac{1}{f_{1}}-1\\
&C^{\left(2\right)}_{iy}=-\frac{A_{i}}{f_{1}}, \,\,
C^{\left(2\right)}_{ii}=\mathcal{C}_{ij}-\frac{1}{f_{1}}\left(A_{i}B_
{j}-
A_{j}B_{i} \right)
\end{eqnarray*}
where the harmonic functions are given by
\begin{eqnarray*}
f_{5}=1+ \frac{Q_{5}}{L}\int_{0}^{L}\frac{du}{\left|x-F
\right|^{2}}\\
f_{1}=1+ \frac{Q_{5}}{L}\int_{0}^{L}\frac{\dot{\left|F
\right|}^{2}du}{\left|x-F \right|^{2}}\\
A_{i}=-\frac{Q_{5}}{L} \int_{0}^{L}du\frac{F_{i,u}}{\left|x-F
\right|^{2}}
\end{eqnarray*}
and the fields $B_{i}$ and $\mathcal{C}_{ij}$ are obtained by
solving
\begin{eqnarray*}
d\mathcal{C}&=&- \ast_{4}df_{5}\\
dB&=&- \ast_{4}dA
\end{eqnarray*}
where the Hodge duality is meant for the the flat space spanned by
$x^{i}$.In our calculations we have used the solution for $B_{i}$
given by
\begin{eqnarray*}
B_{1}&=&B_{2}=0\\
B_{3}&=&- \frac{x_{4}}{x_{3}^{2}+x_{4}^{2}} Z\\
B_{4}&=& \frac{x_{3}}{x_{3}^{2}+x_{4}^{2}} Z\\
Z&=&\frac{Q_{5}}{L}\int_{0}^{L}du \frac{\left(x_{2}-F_{2}
\right)\partial_{u}F_{1}-\left(x_{1}-F_{1}
\right)\partial_{u}F_{2}}{\left|x-F \right|^{2}}.
\end{eqnarray*}
The one brane charge is given by
\begin{eqnarray*}
Q_{1}&=&\frac{Q_{1}}{L}\int_{0}^{L}\dot{\left|F \right|}^{2}du,\,
\text{and}\\
L&=&2\pi Q_{5}.
\end{eqnarray*}
The decoupling limit, which will give $AdS_{3}\times S^{3}$ as the
vacuum, is taken as
\begin{eqnarray*}
f_{5}&=&1+ \frac{Q_{5}}{L}\int_{0}^{L}\frac{du}{\left|x-F
\right|^{2}}\approx \frac{Q_{5}}{L}\int_{0}^{L}\frac{du}{\left|x-F
\right|^{2}}\\
f_{1}&=&1+ \frac{Q_{5}}{L}\int_{0}^{L}\frac{\dot{\left|F
\right|}^{2}du}{\left|x-F \right|^{2}} \approx
\frac{Q_{5}}{L}\int_{0}^{L}\frac{\dot{\left|F
\right|}^{2}du}{\left|x-F \right|^{2}}.
\end{eqnarray*}
We will be interested in regular \cite{LMM} configurations with $F_{3}
=F_{4}=0$. As it was described in \cite{LMM} we expect to get the two
chiral primaries coming from the combination of the metric and the
selfdual part of the flux for the first one, which we will call $
\sigma$, and the combination of the dilaton and the anti-selfdual
part of the flux for the second one which we will call $s^{5}$.
\par
For our purposes we have found useful the change of variables from $F_
{i}\left(u \right)$ to $\rho\left(\phi\right),\, u\left(\phi\right)$
according to
\begin{eqnarray*}
\rho\left(\phi\right)&=&\sqrt{F_{1}^{2}+F_{2}^{2}}\\
\phi&=&\tan\left(\frac{F_{2}\left(u\right)}{F_{2}\left(u\right)}\right).
\end{eqnarray*}
The reader may find some details of the perturbative calculation in
the appendix where we also repeat the treatment of the circular
profile. In the next section we match space time field modes to
fourier modes of the parametrization $\rho,u$.
\par
In order to see the chiral primary $s^{5}$ we have to look at
either the dilaton or the anti-self dual part of the RR field $H$.
Choosing the first option we see that at the first nontrivial order
\begin{eqnarray*}
e^{2\Phi}&\approx&
\frac{f^{0}_{1}}{f^{0}_{5}}+\frac{1}{f^{0}_{5}}\delta
f_{1}-\frac{f^{0}_{1}}{{f_{5}^{0}}^{2}} \delta f_{5}\\
&=&\frac{Q_{1}}{Q_{5}}\left[1+\frac{2\left(r^{2}+\cos^{2}\theta
\right)}{2\pi}\int_{0}^{2\pi} d\tilde{\phi}
\frac{\rho_{0}^{-1}\delta\rho - \delta
u_{,\tilde{\phi}}/Q_{5}}{d^{2}} \right]
\end{eqnarray*}
\par
>From the above we see that the combination that is relevant to the
chiral primary $s^{5}$ is
\begin{equation*}
\delta b= \delta\rho - \frac{\rho_{0}}{Q_{5}}\delta
u_{,\tilde{\phi}}= \sum_{n} b_{n}e^{in\phi},\, b_{-n}=b_{n}^{\dag}
\end{equation*}
The above expansion gives us the fluctuation of the dilaton
\begin{equation*}
\delta\Phi= \sum_{n} b_{n}
\left(\frac{\sin\theta}{\sqrt{r^{2}+1}}\right)^{n}e^{in\phi}.
\end{equation*}
We can read off the field $s^{5}$ from the known expression
\cite{Sezgin}
\begin{equation*}
\delta\Phi\left(t,r,\theta,\phi\right)=2 \sum_{n}
\left|n\right|s^{5}_{n}\left(t,r\right) \sin^{n}\theta e^{in\phi}
\end{equation*}
where we have only included only the relevant for us highest and
lowest weight spherical harmonics in the sum.
\par
In order to extract the chiral primary $\sigma$ we should be looking
at either the metric of the self dual part of the field strength of
the RR field $C^{\left(2\right)}$. Choosing the first way we want to
extract the peturbed metric on the sphere and bring it to the
Lorentz- De Donder gauge. In this gauge the relevant degree of
freedom just scales the volume of the sphere. Expanding the metric
around the ground state and looking at the sphere components we have
that
\begin{eqnarray*}
\frac{d\tilde{s}^{2}_{S^{3}}}{Q_{1}Q_{5}}&=& -\frac{\delta h}{2h_{0}^
{\frac{3}{2}}}\left({A_{\phi}^{0}}^{2}d\phi^{2} +{B_{\psi}^{0}}^{2}d
\psi^{2} \right) +\frac{\delta h}{2Q_{1}Q_{5} h_{0}^{\frac{1}{2}}} d
\mathbf{x}^{2} \\
&&- \frac{1}{\sqrt{h_{0}}} \left(2A^{0}_{\phi}\delta A_{\phi}
d\phi^{2} +2 A_{\phi}^{0}\delta A_{\theta}d\theta d\phi \right) +
\frac{2B_{\psi}^{0}}{\sqrt{h_{0}}}\delta B_{\psi} d\psi^{2}
\end{eqnarray*}
where we have set $h=f_{1}f_{5}$. After putting everything together
we have that
\begin{eqnarray*}
\delta g_{\theta\theta}&=&
\frac{r^{2}+\cos^{2}\theta}{2\sqrt{h_{0}}} \delta h=
\frac{r^{4}-\cos^{4}\theta}{2\pi}\int_{0}^{2\pi}d\tilde{\phi}
\frac{\delta\rho}{d^{4}}\\
&=&\frac{r^{2}-\cos^{2}\theta}{r^{2}+\cos^{2}\theta} \sum_{n} a_{n}
e^{in\phi} \left(\frac{\sin\theta}{\sqrt{r^{2}+1}} \right)^{n}
\left[\left|n\right| + \frac{r^{2}+1+\sin^{2}\theta}
{r^{2}+\cos^{2}\theta} \right]\\
\delta
g_{\psi\psi}&=&\cos^{2}\theta\left[\frac{r^{2}-\cos^{2}\theta}{2\sqrt
{h_{0}}}\delta
h +2\delta B_{\psi} \right] \\
&=&\frac{r^{2}+\sin^{2}\theta+1}{2\pi}\int_{0}^{2\pi}
d\tilde{\phi}\frac{\delta\rho}{d^{4}}
-\frac{2\sqrt{r^{2}+1}\sin\theta}{2\pi}\int_{0}^{2\pi}d\tilde{\phi}
\frac{\sin\left(\tilde{\phi}-\phi\right)}
{d^{2}} \delta\rho_{,\tilde{\phi}}\\
&=&\cos^{2}\theta \sum_{n}a_{n}e^{in\phi}
\left(\frac{\sin\theta}{\sqrt{r^{2}+1}}
\right)^{n}\left[\left|n\right|+
\frac{r^{2}+\sin^{2}\theta+1}{r^{2}+ \cos^{2}\theta} \right].
\end{eqnarray*}
Some of the details of this calculation along with some notation
convetions are explained in the appendix. After performing the gauge
transformation
\begin{equation*}
\theta \rightarrow \theta+ \frac{\sin\theta
\cos\theta}{r^{2}+\cos^{2}\theta}\sum_{n} a_{n} e^{in\phi}
\left(\frac{\sin\theta}{\sqrt{r^{2}+1}} \right)^{n}
\end{equation*}
the metric components of the sphere are scaled as
\begin{eqnarray*}
\delta g_{mn}\left(t,r,\theta,\phi
\right)&=&\bar{g}_{mn}\left(\theta\right)
\sum_{n}\left(\left|n\right|+1 \right) a_{n} e^{in\phi}
\left(\frac{\sin\theta}{\sqrt{r^{2}+1}}
\right)^{n}\\
&=&2\bar{g}_{mn}\left(\theta\right) \sum_{n}\left|n\right|
\sigma_{n}\left(t,r\right)\sin^{n}\theta e^{in\phi}
\end{eqnarray*}
where $\bar{g}_{mn}$ is metric on $S^{3}$.
\par
The previous considerations helped us translate degrees of freedom
of the curve to spacetime fields of minimal six dimensional SUGRA
coupled to one tensor multiplet. The correctly normalized action for
the above chiral primaries is given by
\begin{eqnarray*}
S=&&\sum_{n}\frac{2\left|n\right|}{\left|n\right|+1} \int_{AdS_{3}}
dx^{3} \left[\left(\left|n\right|+1\right)\left( \left(\nabla
s^{5}_{n}\right)^{2}+
\left|n\right|\left(\left|n\right|-2\right)\left(s^{5}_{n}\right)^{2}
\right)\right]\\
+&&\sum_{n}\frac{2\left|n\right|}{\left|n\right|+1} \int_{AdS_{3}}
dx^{3} \left[\left(\left|n\right|-1\right)\left(
\left(\nabla\sigma_{n}\right)^{2}+
\left|n\right|\left(\left|n\right|-2\right)\left(\sigma_{n}\right)^{2}
\right)\right]
\end{eqnarray*}
After restricting the system to the supersymmetric case we have the
canonical expansion of the fields
\begin{eqnarray*}
s^{5}&=&\sum_{n} s^{5}_{n} Y^{n}= \frac{1}{2} \sum_{n}
\frac{1}{\sqrt{2\left|n\right|}}e^{in\phi}
\left(\frac{\sin\theta}{\sqrt{r^{2}+1}} \right)^{n} c_{n}\\
\sigma&=& \sum_{n} \sigma_{n} Y^{n}= \frac{1}{2} \sum_{n}
\sqrt{\frac{1}{2\left|n\right|}\frac{\left|n\right|+1}{\left|n
\right|-1}}e^{in\phi}
\left(\frac{\sin\theta}{\sqrt{r^{2}+1}} \right)^{n} d_{n}.
\end{eqnarray*}
Comparing the components of the SUGRA fields that we previously
calculated we see that the identification is
\begin{eqnarray*}
b_{n}=\sqrt{\frac{\left|n\right|}{2}}c_{n}\\
a_{n}=\sqrt{\frac{1}{2}\frac{\left|n\right|}{n^{2}-1}}d_{n}
\end{eqnarray*}
\par
The angular momentum of the full solution is given by
\begin{eqnarray*}
J&=&J_{L}+J_{R}=\frac{1}{2\pi} \int_{0}^{L}du
\left(F_{1}\left(u\right) \dot{F}_{2}\left(u\right)-
\dot{F}_{1}\left(u\right) F_{2}\left(u\right)\right)-Q_{1}Q_{5}\\
&=&\frac{1}{2\pi}\int_{0}^{2\pi}d\phi
\rho^{2}\left(\phi\right)-Q_{1}Q_{5}
\end{eqnarray*}
At this point we would like to separate the zero mode dependence of
$\rho\left(\phi\right)$ as
\begin{eqnarray*}
&\rho\left(\phi\right)=\rho_{0}+\delta\rho\left(\phi\right)\\
&\int_{0}^{2\pi}d\phi\,\delta\rho\left(\phi\right)=0.
\end{eqnarray*}
The result for the angular momentum is
\begin{eqnarray*}
J=\rho_{0}^{2}+ \sum_{n>1}a_{n}a_{-n}-Q_{1}Q_{5}
\end{eqnarray*}
For the case of the $D1-D5$ system we need to keep $Q_{1}Q_{5}$
fixed, which gives the constrain
\begin{equation*}
N=\frac{Q_{5}^{2}}{L} \int_{0}^{L}du \dot{\left|F\right|}^{2}=
\frac{Q_{5}}{2\pi}\int_{0}^{2\pi}d\phi\,
\frac{\rho_{,\phi}^{2}+\rho^{2}}{u_{,\phi}}
\end{equation*}
We can solve the above constrain for $\rho_{0}$ perturbatively in
fluctuations of $\rho$ and $u$.
\begin{equation*}
\rho_{0}^{2}\approx N+ \frac{1}{2\pi}
\int_{0}^{2\pi}d\phi\,\delta\rho
\delta\rho_{,\phi\phi}-\frac{1}{2\pi}
\int_{0}^{2\pi}d\phi\,\left(\delta\rho -
\frac{\rho_{0}}{Q_{5}}\delta u_{,\tilde{\phi}} \right)^{2}
\end{equation*}
Using the expression that we previously found for the angular
momentum we find
\begin{eqnarray*}
J&=&-\frac{1}{2\pi} \int_{0}^{2\pi}d\phi\,\left(\delta\rho -
\frac{\rho_{0}}{Q_{5}}\delta u_{,\tilde{\phi}} \right)^{2} +
\frac{1}{2\pi} \int_{0}^{2\pi}d\phi\,\left(\delta\rho
\delta\rho_{,\phi\phi}+\delta\rho^{2}\right)\\
&=& -\sum_{n>0} b_{n}b_{-n}-\sum_{n>1} \left(n^{2}-1 \right)a_{n}a_{-
n}\\
&=& -\sum_{n>0} n c_{n}c_{-n}-\sum_{n>1} n d_{n}d_{-n}
\end{eqnarray*}

\section{Non-linear analysis}
The  f1f5  solution provids a general non-singular 1/2 BPS solution
of 6D Sugra .It is parametrized by the string-like coordinates F
whose dynamics is of interest. One can think of this in analagy with
extended (soliton-like) solutions of nonlinear field
theory\cite{GJS}. In this spirint one would like to establish the
dynamics of general 1/2 BPS configuration in Supergravity by direct
evaluation of the action.Some aspects of the result could be infered
from the extended string-like (or superube) nature othe solution.
The direct Sugra confirmation is nevertheless of importance,
especialy regarding the question of correlation  and interactions in
the theory.
 With this purpose we begin first by direct  evaluation the symplectic form
from Sugra. In this  we will follow the methods of \cite{MR}
which are based on the covariant ZCW symplectic form \cite{CW,Zuckerman:1989cx}.
According to
this method which is generaly applicable to every field theory described by a
Lagrangian density $L\left(\phi_{i},\partial_{l}\phi_i\right)$,
with fields and their first derivatives, the symplectic form
is given by
\begin{equation*}
J^{l}=\frac{\delta L\left(\phi_{i},\partial_{l}\phi_i\right)}{\delta
\partial_{l}\phi_{i}}\wedge \delta\phi_{i}.
\end{equation*}
As it was pointed out in \cite{MR}. while using the above expression
one has to use a regular gauge choice for both the fields $\phi_{i}$
and the variations $\delta\phi_{i}$. The effect of gauge
transformations is the addition of total derivative terms in the
vector symplectic density. In the case where one has well behaved
transformations these total derivative terms can be dropped. In the
case where one wants to correct a singular gauge choice for the filed
variations $\delta\phi_{i}$ these total derivative terms cannot be
dropped \cite{MR}.
\par
For the space of solutions $F_{3}=F_{4}=0$ that we are interested in
we will use the dimensionally reduced 6d Einstein frame action
\begin{eqnarray*}
S=\int dx^{6}\sqrt{-g}\left[R-\frac{1}{12}
e^{\sqrt{2}a}H^{2}-\frac{1}{2} \left(\nabla a\right)^{2}\right]=\\
\int dx^{6}\sqrt{-g}\left[g^{mn}\left(\Gamma_{ml}^{a}
\Gamma_{an}^{l}-\Gamma_{mn}^{a}\Gamma_{la}^{l} \right)-\frac{1}{12}
e^{\sqrt{2}a}H^{2}-\frac{1}{2} \left(\nabla a\right)^{2}\right]
\end{eqnarray*}
in order to evaluate the symplectic form.
\par
In the above action we have dropped the total derivative terms
that come from partial integrations necessary to reach the first
derivative form of the action. One could check that such a total
derivative term \cite{Wald} gives
no contribution. The $D1-D5$ solution in the ten dimensional string
frame is given by
\begin{eqnarray*}
&d\tilde{s}^{2}=\frac{1}{\sqrt{f_{1}f_{5}}}\left[-\left(dt-
A_{i}dx^{i} \right)^{2} + \left(dy +B_{i}dx^{i} \right)^{2} \right]+
\sqrt{f_{1}f_{5}}
d\bold{x}^{2}+ \sqrt{\frac{f_{1}}{f_{5}}} d\mathbf{z}^{2}\\
&e^{2\Phi}=\frac{f_{1}}{f_{5}}\\
&C^{\left(2\right)}_{ti}=\frac{B_{i}}{f_{1}},\,\,
C^{\left(2\right)}_{ty}=\frac{1}{f_{1}}-1\\
&C^{\left(2\right)}_{iy}=-\frac{A_{i}}{f_{1}}, \,\,
C^{\left(2\right)}_{ij}=\mathcal{C}_{ij}-\frac{1}{f_{1}}\left(A_{i}B_
{j}-
A_{j}B_{i} \right).
\end{eqnarray*}
The harmonic functions that appear in the above equations are
\begin{eqnarray*}
f_{5}=1+ \frac{Q_{5}}{L}\int_{0}^{L}\frac{du}{\left|x-F
\right|^{2}}\\
f_{1}=1+ \frac{Q_{5}}{L}\int_{0}^{L}\frac{\dot{\left|F
\right|}^{2}du}{\left|x-F \right|^{2}}\\
A_{i}=-\frac{Q_{5}}{L} \int_{0}^{L}du\frac{F_{i,u}}{\left|x-F
\right|^{2}}
\end{eqnarray*}
and the fields $B_{i}$ and $\mathcal{C}_{ij}$ are obtained by
solving
\begin{eqnarray*}
d\mathcal{C}&=&- \ast_{4}df_{5}\\
dB&=&- \ast_{4}dA
\end{eqnarray*}
where the Hodge duality is meant for the flat space spanned by
$x^{i}$. In order to transform the solution to the ten dimensional
Einstein frame we need to scale the metric as
\begin{equation*}
g_{\mu\nu}=e^{\frac{\phi}{2}}\tilde{g}_{\mu\nu}.
\end{equation*}
It is true that the above transformation generates total derivative
terms. The behavior of the dilaton near the curve
$\mathbf{x}=\mathbf{F}\left(u \right)$ is such that possible
contributions are zero for our case. Going back to the six
dimensional Einstein frame we have the ansatz
\begin{eqnarray*}
ds^{2}&=& -\frac{1}{\sqrt{f_{1}f_{5}}}\left(dt- A_{i}dx^{i}
\right)^{2} + \left(dy +B_{i}dx^{i} \right)^{2} + \sqrt{f_{1}f_{5}}
d\bold{x}^{2}\\
a&=&\sqrt{2}\phi
\end{eqnarray*}
The scalar $a$ comes from a combination of the ten dimensional
dilaton with the volume of the torus. We will need the solution for
$B_{i}$ which is given by
\begin{eqnarray*}
B_{1}&=&B_{2}=0\\
B_{3}&=&- \frac{x_{4}}{x_{3}^{2}+x_{4}^{2}} Z\\
B_{4}&=& \frac{x_{3}}{x_{3}^{2}+x_{4}^{2}} Z\\
Z&=&\frac{Q_{5}}{L}\int_{0}^{L}du \frac{\left(x_{2}-F_{2}
\right)\partial_{u}F_{1}-\left(x_{1}-F_{1}
\right)\partial_{u}F_{2}}{\left|x-F \right|^{2}}
\end{eqnarray*}
We need to observe that in this gauge choice
\begin{eqnarray*}
\nabla A=\nabla B=A\cdot B=0.
\end{eqnarray*}
\par
Fixing our Cauchy surface by $l=t$ we have
\begin{equation*}
J^{t}=J_{G}^{t}+J_{H}^{t}+J_{a}^{t}+J_{bdry}^{t}
\end{equation*}
where
\begin{eqnarray*}
J_{G}^{l}&=&-\delta\Gamma_{mn}^{l}\wedge\delta\left(\sqrt{-g}g^{mn}
\right)
+\delta\Gamma_{mn}^{n}\wedge\delta\left(\sqrt{-g}g^{lm}\right)\\
J_{H}^{l}&=&-\frac{1}{2} \delta\left(\sqrt{-g}e^{\sqrt{2}a} H^{lmn}
\right)\wedge\delta\left(C_{mn} \right)\\
J_{a}^{l}&=&- \delta\left(\sqrt{-g}g^{lm}\partial_{m}a \right)\wedge
\delta\left(a\right)
\end{eqnarray*}
and $J_{bdry}^{t}$ is the contribution that comes from the gauge
transformation which is necessary to show regularity of the metric
$\delta g_{mn}$ and the gauge field $\delta C_{mn}$. We will analyze
this term separately. As we show in the appendix the bulk
contribution is given by
\begin{equation*}
J_{bulk}^{t}=-\int d^{4}x \partial_{d} \left\{\frac{1}{2}
\epsilon_{ijad} \delta\left[\frac{A^{a}}{f_{5}}\right]\wedge \delta
\left(C_{ij}\right)\right\}.
\end{equation*}
The above total integral receives contribution from a surface of $S^
{1}\times S^{2}$ topology which describes a "tube" surrounding the
ring at $\mathbf{x}=\mathbf{F}\left(u\right)$. As we show in the
appendix the value of this integral is given by
\begin{eqnarray*}
\int d^{4}x \partial_{d}\left[ \frac{1}{2} \epsilon_{ijad}
\delta\left[\frac{A^{a}}{f_{5}}\right]\wedge \delta
\left(C_{ij}\right)\right]=- 2\pi \int_{0}^{L} du \frac{1}{\left|
\dot{F}\right|^{2}}\left(\dot{F}_{1} \delta \dot{F}_{1} +
\dot{F}_{2} \delta \dot{F}_{2} \right)\wedge \left(\dot{F}_{1}
\delta F_{1} + \dot{F}_{2} \delta F_{2}  \right).
\end{eqnarray*}

\subsection{Contribution coming from regularity gauge transformation}
In order to show regularity of the fields under variation we need to
perform gauge transformations for both the coordinates and the two
form gauge field. For the present case we find that the only
transformation that generates some finite contribution is the
general coordinate transformation
\begin{equation*}
x^{i}\rightarrow x^{i}+ \xi^{i}\left(x^{j},\delta F_{j}, \delta
\dot{F}_{j} \right)
\end{equation*}
that we have to perform in order to show regularity of the metric
$g_{mn}+\delta g_{mn}$. The above transformation generates total
derivative term contribution in the symplectic form \cite{CW}. The
contribution coming from infinity can be discarded since these terms
would be proportional to the variation of the charges that we keep
fixed. The place that we need to specify the coordinate
transformation is near the curve $x_{i}=F_{i}$. The transformation
\begin{eqnarray*}
\delta C_{mn}&\rightarrow& \delta C_{mn}+\xi^{p}H_{pmn}-2
\partial_{\left[m\right.}\left(\xi^{p}C_{\left. n\right]p}\right)\\
\delta g_{mn}&\rightarrow& \delta g_{mn} + \nabla_{\left(m
\right.}\xi_{\left. n\right)}
\end{eqnarray*}
close to the curve is given by
\begin{eqnarray*}
\xi^{i}=\delta F^{i}- \frac{\left(\vec{x}-\vec{F}\right)\cdot
\delta\vec{l}}{\dot{\left|F \right|}^{2}}l^{i}-
\frac{\left(\vec{x}-\vec{F}\right)\cdot \vec{l}}{\dot{\left|F
\right|}^{2}}\delta l^{i}+2 \frac{\left(\vec{x}-\vec{F}\right)\cdot
\vec{l}}{\dot{\left|F \right|}^{4}}  \left(\vec{l}\cdot\delta \vec{l}
\right)l^{i}\\
-\frac{\left(\vec{x}-\vec{F}\right)\cdot \delta\vec{n}}{\dot{\left|F
\right|}^{2}}n^{i}- \frac{\left(\vec{x}-\vec{F}\right)\cdot
\vec{n}}{\dot{\left|F \right|}^{2}}\delta n^{i}+2
\frac{\left(\vec{x}-\vec{F}\right)\cdot \vec{n}}{\dot{\left|F
\right|}^{4}} \left(\vec{n}\cdot\delta \vec{n}
\right)n^{i}\\
\vec{l}\left(u\right)=\left(
                    \begin{array}{cccc}
                    \dot{F_{1}}\left(u\right),
             &\dot{F}_{2}\left(u\right),
             &0, & 0 \\
                    \end{array}
                  \right)\\
\vec{n}\left(u\right)=\left(
                    \begin{array}{cccc}
                    \dot{F_{2}}\left(u\right),
             &-\dot{F}_{1}\left(u\right),
             &0, & 0 \\
                    \end{array}
                  \right)
\end{eqnarray*}
The vector $\vec{l}$ is tangent to the curve $\vec{F}$ and the
vector $\vec{n}$ is one of its normal vectors. A way to see this is,
is to say that $g_{mn}+\delta g_{mn}$ is still in the family of
solutions that we consider with $\mathbf{F}\rightarrow
\mathbf{F}+\delta\mathbf{F}$. In order to show regularity we would
have to follow steps similar to \cite{LMM}. Starting from the background
metric $g_{mn}$ we would need to perform the above mentioned
coordinate transformation in order to show that the total metric
$g_{mn}+\delta g_{mn}$ is regular.
\par
Unlike the case of the gauge field gauge transformations, the coordinate transformation that we need to perform induces total derivative terms quadratic in the gauge transformation parameters $\xi^{i}$. The quadratic terms in the symplectic form can be obtained from the linear ones simply by substituting $\delta g_{mn}\rightarrow \nabla_{\left(m\right.}\xi_{\left.n\right)}$. The above argument holds since a pure gauge satisfies the linearized equations of motion.
\par
We observe that $H^{tmn}$ go to zero fast enough close to the curve
not to give any contribution from the three form term in the action.
The only terms that contribute coming from the gravitational total
derivative \cite{CW} are
\begin{eqnarray*}
\Delta J^{t}=- \sqrt{-g}\, \nabla_{n}\left[c^{n}\left(\delta g_{m,n},\xi^{l} \right)+ \frac{1}{2}c^{n}\left(\nabla_{\left(m\right.}\xi_{\left. n\right)} ,\xi^{l} \right) \right]
\end{eqnarray*}
\begin{eqnarray*}
c^{n}\left(\delta g_{m,n},\xi^{l} \right)= \left(\nabla_{m}\delta g^{mn}+\nabla^{n} \delta \ln g \right)\wedge \xi^{t}+ \nabla^{n}\delta g^{mt}\wedge \xi_{m} \\
+\nabla_{m}\xi^{t}\wedge \delta g^{mn}+\frac{1}{2} \nabla^{n}\xi^{t}\wedge\delta\ln g - \left(t\leftrightarrow n\right)
\end{eqnarray*}
\par
As we see in the appendix after integrating the above expression and
taking the appropriate limit we have
\begin{equation*}
\Delta J^{t}=2\pi \int_{0}^{L} du \frac{1}{\left|
\dot{F}\right|^{2}}\left(\dot{F}_{2} \delta \dot{F}_{1} -\dot{F}_{1}
\delta \dot{F}_{2} \right)\wedge \left(\dot{F}_{2} \delta F_{1} -
\dot{F}_{1} \delta F_{2}  \right)
\end{equation*}
\par
After we add the contribution that comes from the total derivative
term we have the final form \footnote{The expected form for this
symplectic form was described in Rychkov,talk at Strings 2005, also \cite{Rychkov:2005ji}}
\begin{equation*}
J^{t}=2\pi \int_{0}^{L}\,du \delta\dot{F}_{i}\wedge \delta F_{i}.
\end{equation*}

\subsection{Non-linear dynamics}
In this section we would like to consider the full dynamics of the system. For simplicity we will concentrate on two of the degrees of freedom. The analysis in general will be analogous. One can write down an action consisting of the symplectic form (derived in the previous section) plus the angular momentum of the NS sector.
\begin{equation*}
S= \int dt\, du \left[\dot{F}_{1}F_{1}^{\prime}+\dot{F}_{2}F_{2}^
{\prime} +F_{1}F_{2}^{\prime}-F_{2}F_{1}^{\prime}-{F_{1}^{\prime}}^
{2}-{F_{2}^{\prime}}^{2} \right]
\end{equation*}
Our goal is to relate this dynamical system to the collective dynamics given in sect. 2. We will demonstrate agreemnt by a change of variables
\begin{eqnarray*}
F_{1}\left(t,u\right)&=&r\left(t,u\right) \cos\left(\phi\left(t,u
\right)\right)\\
F_{2}\left(t,u\right)&=&r\left(t,u\right) \sin\left(\phi\left(t,u
\right)\right).
\end{eqnarray*}
The action now takes the form
\begin{equation*}
S=\int dt\, du \left[\dot{r}r^{\prime}+ \dot{\phi}\phi^{\prime}r^{2}+
\phi^{\prime}r^{2}- {r^{\prime}}^{2}-r^{2}{\phi^{\prime}}^{2} \right]
\end{equation*}
while the non-trivial constrain now reads
\begin{equation*}
\int_{0}^{2\pi}\left[r^{2}{\phi^{\prime}}^{2}+{r^{\prime}}^{2} \right]
du= 2\pi Q_{1}Q_{5}.
\end{equation*}
We may shift the field $\phi$ by a classical background
\begin{eqnarray*}
\phi &\rightarrow& u+\tilde{\phi}\\
r &\rightarrow& r_{0}+ \tilde{r}
\end{eqnarray*}
and also split the the field into a zero mode $r_{0}$ and its
fluctuating part which we denote by $\tilde{r}$. After doing so one
can show, by non-linear field redefinitions, that up to cubic order
in fluctuations the system is described by the action
\begin{equation*}
S= \int dt du \left[\hat{D}\dot{r}\int^{u}du^{\prime}\hat{D}r+ \dot{z}
\int^{u}du^{\prime}z+ \hat{D}r\hat{D}r - z^{2}\right]- 2\pi \int dt\,
r_{0}^{2}
\end{equation*}
where
\begin{eqnarray*}
\hat{D}&=&i\partial_{u}+1\\
z&=&r_{0}\phi^{\prime}+\tilde{r}.
\end{eqnarray*}
In order to make contact with the fermion droplet in the action angle
representation we redefine
\begin{equation*}
\rho^{2}=\hat{D}r.
\end{equation*}
As we have seen in the perturbative analysis the field $\rho^{2}$ is
directly related to the gravity chiral primary field $\sigma$ and $z$
is related to the field $\sigma^{10}$. The action now reads,
\begin{equation*}
S= \int dt du \left[\dot{\rho}^{2}\int^{u}du^{\prime}\rho^{2}+ \dot{z}
\int^{u}du^{\prime}z+ \rho^{4} - z^{2}\right]- 2\pi \int dt\, \rho_{0}
^{4}.
\end{equation*}
In order to make contact with the non-linear chiral boson we perform
the field dependent coordinate transformation
\begin{eqnarray*}
x&=&\rho \cos u\\
y_{\pm}&=&\rho \sin u
\end{eqnarray*}
and the action now reads
\begin{eqnarray*}
S=&&\int dt\,dx\,\left[4\dot{y}_{+} \int^{x}dx^{\prime}{y}_{+} - 4\dot
{y}_{-} \int^{x}dx^{\prime}{y}_{-}-\frac{4}{3}\left(y^{3}_{+}-y^{3}_
{-} \right)-4x^{2}\left(y_{+}-y_{-} \right)  \right]\\
+&&\int dt\,dx\,\frac{x\partial_{x}y_{+}-y_{+}}{y_{+}^{2}+x^{2}}\left
[\dot{z}_{+}-\frac{x\dot{y}_{+}}{x\partial_{x}y_{+}-y_{+}}\partial_{x}
z_{+} \right]\int^{x}dx^{\prime}\frac{x\partial_{x}y_{+}-y_{+}}{y_{+}^
{2}+x^{2}}z_{+}\\
-&&\int dt\,dx\,\frac{x\partial_{x}y_{-}-y_{-}}{y_{-}^{2}+x^{2}}\left
[\dot{z}_{-}-\frac{x\dot{y}_{-}}{x\partial_{x}y_{-}-y_{-}}\partial_{x}
z_{-} \right]\int^{x}dx^{\prime}\frac{x\partial_{x}y_{-}-y_{-}}{y_{-}^
{2}+x^{2}}z_{-}\\
-&&\int dt\,dx \left[\frac{y_{+}-x\partial_{x}y_{+}}{y_{+}^{2}+x^{2}}
z_{+}^{2}- \frac{y_{-}-x\partial_{x}y_{-}}{y_{-}^{2}+x^{2}}z_{-}^{2}
\right]- 2\pi \int dt\, \rho_{0}^{4}.
\end{eqnarray*}
At this point we shift by the background
\begin{eqnarray*}
y^{0}_{\pm}&=&\pm\sqrt{\rho_{0}^{2}-x^{2}}=\pm \phi^{0}\\
\rho_{0}^{2}&\approx&\sqrt{Q_{1}Q_{5}}
\end{eqnarray*}
which corresponds to the $AdS_{3}\times S^{3}$ background in the
decoupling limit of the $D1-D5$ system. We also need to perfom the
shifts
\begin{eqnarray*}
\pm z_{\pm}&\rightarrow& \frac{1}{2\rho_{0}^{2}}\phi^{0} \partial_{x}
\left[\phi^{0}\partial_{x}\left(xy_{\pm}\right)\int^{x}\frac{dx^
{\prime}}{\phi^{0}}z_{\pm} \right]+ \frac{1}{2\rho_{0}^{2}}\phi^{0}
\partial_{x}\left(xy_{\pm}\right)z_{\pm}\\
\pm y_{\pm}&\rightarrow&-\frac{1}{8\rho_{0}^{2}}
\partial_{x}\left[x\partial_{x}z_{\pm}
\int^{x}\frac{dx^{\prime}}{\phi^{0}}z_{\pm} \right]
\end{eqnarray*}
so that we will have the Poisson brackets
\begin{eqnarray*}
\left\{y_{\pm}\left(x\right),y_{\pm}\left(x^{\prime}\right) \right\}
&=&\pm 4 \partial_{x}\delta \left(x-x^{\prime}\right)\\
\left\{z_{\pm}\left(x\right),z_{\pm}\left(x^{\prime}\right)
\right\}&=&\pm \phi^{0}\left(x\right)
\partial_{x}\left[\phi^{0}\left(x\right) \delta
\left(x-x^{\prime}\right)\right].
\end{eqnarray*}
The resulting Hamiltonian is given by
\begin{eqnarray*}
H=&&\int dt\,dx\left[ 4\phi^{0} \left(y^{2}_{+}+y^{2}_{-}\right)+
\frac{1}{q^{0}} \left(z^{2}_{+}+z^{2}_{-}\right) \right]+\frac{4}{3}
\int dt\,dx\left(y^{3}_{+}-y^{3}_{-} \right)\\
+&&\frac{1}{\rho^{2}_{0}}\int dt\,dx\left[z_{+}\partial_{x}\left
[\partial_{x}\left(xy_{+}\right)\int^{x}\frac{dx^{\prime}}{\phi^{0}}z_
{+} \right]-\phi^{0}y_{+}\partial_{x}\left[x\partial_{x}z_{+} \int^{x}
\frac{dx^{\prime}}{\phi^{0}}z_{+} \right] \right]\\
-&&\frac{1}{\rho^{2}_{0}}\int
dt\,dx\left[z_{-}\partial_{x}\left[\partial_{x}\left(xy_{-}\right)
\int^{x}\frac{dx^{\prime}}{\phi^{0}}z_{-}
\right]-\phi^{0}y_{-}\partial_{x}\left[x\partial_{x}z_{-}
\int^{x}\frac{dx^{\prime}}{\phi^{0}}z_{-} \right] \right].
\end{eqnarray*}
\par
The structure of the above Hamiltonian is of the form
\begin{equation*}
H=H\left(y\right)+ H_{int}
\end{equation*}
where $HH\left(y\right)$ is the cubic collective Hamiltonian of the
free fermions and
\begin{equation*}
H_{int}=y_{+}T\left(z_{+}\right)+ y_{-}T\left(z_{-}\right)
\end{equation*}
has the form of the interacting Hamiltonian given in the first
section.Let us recapitulate what was done in this subsection. Starting with the
linear F-system we performed a change of variables relating the gravitational 1/2 BPS configuration to the interacting picture of $S_{N}$ orbifold collective field theory. The change of variables was
deduced from the linearized analysis (giving the physical degrees of
freedom in sect.3 ) and the change from angular to cartesian coordinates
known in the case of a pure fermion droplet. The existence of such a change of variables relating the linear with the nonlinear version of our theory is associated with the integrable property of the system. In the CFT or matrix model version this integrable feature was already
identified in ref.16,it can also be interpreted as coming from the
integrability of the matrix-vector model described in Sect.2.

\section{Conclusions}

In this paper we have considered in some detail the hamiltonian
descriptions of BPS states in AdS3 xS3 Sugra . Elaborating on the
construction of [16] we have given a summary of the collective
field description which was associated with the correlators of the SN orbifold conformal field theory. We have identified a simple set of
elementary coordinates responsible for the dynamics of the BPS sector of the theory. The system consists of a set of eigenvalu coordinates plus
the fermionic zero modes of the compactifying manifold X .As such it
extends the fermion droplet of the D3 brane theory ,here one has a fermion droplet interacting with boundary degrees of freedom.
Direct calculations in AdS3xS3xT4 Sugra were performed to exibit the analogous structure from supergravity. To this end the evaluation of the
symplectic form for the general BPS configuratin in Sugra is given. After an identification of physical degrees of freedom a comparison with the collective field picture is accomplished. This involved a transformation from the action-angle variables desribing the general D1D5 2-charge configuration to a physical set of variables.
Further studies of this system are clearly of interest.This involves
a correspondence with the 'supertube' picture and the question of
its quantum mechanical description. Extension to more general
3-charge configurations are of further major interest as they
offer a promise for studies of black hole configurations in
this approach.

\section*{Acknowledgments}
We would like to thank Jeffrey Murugan and Robert McNees for useful
discussions. We also thank V. S. Rychkov for communication, especially for the quadratic terms in the general coordinate contribution.

\newpage
\appendix

\section{Appendix A}

\subsection{$AdS_{3}\times S^{3}$}
The way that we may see $AdS_{3}\times S^{3}$ is by setting
\begin{eqnarray*}
F_{1}&=& \sqrt{Q_{1}Q_{5}}\cos\left(\frac{u}{Q_{5}}\right),\, F_{2}=
\sqrt{Q_{1}Q_{5}}\sin\left(\frac{u}{Q_{5}}\right)\\
x_{1}&=&\sqrt{Q_{1}Q_{5}} \sqrt{r^{2}+1}\sin\theta \cos\phi\\
x_{2}&=&\sqrt{Q_{1}Q_{5}} \sqrt{r^{2}+1}\sin\theta \sin\phi\\
x_{3}&=&\sqrt{Q_{1}Q_{5}} r \cos\theta \cos\psi\\
x_{3}&=&\sqrt{Q_{1}Q_{5}} r \cos\theta \sin\psi.
\end{eqnarray*}
Plugging the above to the solution we find the harmonic functions
\begin{eqnarray*}
f_{1,5}^{0}=\frac{Q_{5,1}^{-1}}{r^{2}+\cos^{2}\theta}\\
A_{\phi}^{0}=\frac{\sin^{2}\theta}{r^{2}+\cos^{2}\theta}\\
B_{\psi}^{0}=\frac{\cos^{2}\theta}{r^{2}+\cos^{2}\theta}
\end{eqnarray*}
which yield the six dimensional Einstein frame metric
\begin{eqnarray*}
\frac{ds_{6}^{2}}{\sqrt{Q_{1}Q_{5}}}=\left(r^{2}+\cos^{2}\theta
\right)\left[-\left(dt- \frac{\sin^{2}\theta
d\phi}{r^{2}+\cos^{2}\theta} \right)^{2}+
  \left(d\chi+
\frac{\cos^{2}\theta d\psi}{r^{2}+\cos^{2}\theta}
\right)^{2}\right]\\+ \frac{1}{r^{2}+\cos^{2}\theta} \left[\left(
r^{2}+\cos^{2}\theta\right)\left(\frac{dr^2}{r^{2}+1}+d\theta^{2}
\right)+r^{2}\cos^{2}\theta d\phi^{2}+ \left(r^{2}+1\right)
\sin^{2}\theta d\phi^{2} \right].
\end{eqnarray*}
After perfoming the change of coordinates
\begin{equation*}
\tilde{\phi}=\phi-t,\,\, \tilde{\psi}=\psi+y
\end{equation*}
we find the $AdS_{3}\times S^{3}$ metric in global
coordinates
\begin{eqnarray*}
\frac{ds_{6}^{2}}{\sqrt{Q_{1}Q_{5}}} = -\left(r^{2}+1\right)dt^{2}+
\frac{dr^2}{r^{2}+1} + r^{2} dy^{2}+ d\theta^{2}+ \sin^{2}\theta
d\phi^{2}+ \cos^{2}\theta d\psi^{2}.
\end{eqnarray*}

\subsection{From $F_{1}\left(u\right),F_{2}\left(u\right)$ to
$\rho\left(\phi\right),u\left(\phi \right)$}

In the next section we correspond modes of the curve to modes of the
two chiral primaries $\sigma_{n}$ and $s^{5}_{n}$ where $n$ is a
quantum number that has to do with the isometry of the background metric
related to the killing vector $\partial_{\phi}$. As we can see from
the coordinate transformation
\begin{eqnarray*}
x_{1}=\sqrt{Q_{1}Q_{5}} \sqrt{r^{2}+1}\sin\theta \cos\phi\\
x_{2}=\sqrt{Q_{1}Q_{5}} \sqrt{r^{2}+1}\sin\theta \sin\phi
\end{eqnarray*}
the angle $\phi$ is related to the angle
$\phi=\arctan\left(\frac{F_{2}\left(u\right)}{F_{1}\left(u\right)}
\right)$. We now perform the change of variables
\begin{eqnarray*}
\rho^{2}\left(\phi\right)&=&
F_{1}^{2}\left(u\left(\phi\right)\right)+
F_{1}^{2}\left(u\left(\phi\right)\right)\\
\phi&=&\arctan\left(\frac{F_{2}\left(u\right)}{F_{1}\left(u\right)}
\right)
\end{eqnarray*}
After inverting the last equation one is able to express $u$ in
terms of $\phi$. In terms of the new variables the functions
calculated from the curve $\rho\left(\phi\right),u\left(\phi
\right)$ read

\begin{eqnarray*}
f_{5}=1+ \frac{1}{2\pi}\int_{0}^{2\pi}d\tilde{\phi}\frac{
u_{,\tilde{\phi}}}{d\left(\rho,\theta, \phi;
\rho\left(\tilde{\phi}\right),\tilde{\phi}\right)^{2}} \approx
\frac{1}{2\pi}\int_{0}^{2\pi}d\tilde{\phi}\frac{
u_{,\tilde{\phi}}}{d\left(\rho,\theta, \phi; \rho\left(\tilde{\phi}
\right),\tilde{\phi}\right)^{2}}\\
f_{1}=1+
\frac{1}{2\pi}\int_{0}^{2\pi}\frac{d\tilde{\phi}}{u_{,\tilde{\phi}}}
\frac{
\rho_{\tilde{\phi}}^{2}+ \rho^{2}} {d\left(\rho,\theta, \phi;
\rho\left(\tilde{\phi}\right),\tilde{\phi}\right)^{2}} \approx
\frac{1}{2\pi}\int_{0}^{2\pi}\frac{d\tilde{\phi}}{u_{,\tilde{\phi}}}
\frac{\rho_{\tilde{\phi}}^{2}+
\rho^{2}} {d\left(\rho,\theta,
\phi;\rho\left(\tilde{\phi}\right),\tilde{\phi}\right)^{2}}
\end{eqnarray*}
\begin{eqnarray*}
A_{\phi}=\frac{\sqrt{Q_{1}Q_{5}}\sqrt{r^{2}+1}\sin\theta}{2\pi}
\int_{0}^{2\pi}d\tilde{\phi}\,\frac{\rho \cos\left(\phi-\tilde{\phi}
\right)+ \sin\left(\tilde{\phi}-\phi
\right)\partial_{\tilde{\phi}}\rho}{d\left(\rho,\theta, \phi;
\rho\left(\tilde{\phi}\right),\tilde{\phi}\right)^{2}}\\
A_{r}=\frac{\sqrt{Q_{1}Q_{5}}r\sin\theta}{2\pi \sqrt{r^{2}+1}}
\int_{0}^{2\pi}d\tilde{\phi}\,\frac{\partial_{\tilde{\phi}}\rho
\cos\left(\phi-\tilde{\phi} \right)+ \rho \sin\left(\phi
-\tilde{\phi}\right)}{d\left(\rho,\theta, \phi;
\rho\left(\tilde{\phi}\right),\tilde{\phi}\right)^{2}}\\
A_{\theta}=\frac{\sqrt{Q_{1}Q_{5}}\sqrt{r^{2}+1}\cos\theta}{2\pi }
\int_{0}^{2\pi}d\tilde{\phi}\,\frac{\partial_{\tilde{\phi}}\rho
\cos\left(\phi-\tilde{\phi} \right)+ \rho \sin\left(\phi
-\tilde{\phi}\right)}{d\left(\rho,\theta, \phi;
\rho\left(\tilde{\phi}\right),\tilde{\phi}\right)^{2}}\\
B_{\psi}= -A_{\phi} - \frac{1}{2\pi}\int_{0}^{2\pi}d\tilde{\phi}
\frac{\rho^{2}}{d\left(\rho,\theta, \phi;
\rho\left(\tilde{\phi}\right),\tilde{\phi}\right)^{2}}
\end{eqnarray*}
where
\begin{equation*}
d\left(\rho,\theta, \phi;
\rho\left(\tilde{\phi}\right),\tilde{\phi}\right)^{2}=
Q_{1}Q_{5}\left(r^{2}+\sin^{2}\theta
\right)+\rho^{2}\left(\tilde{\phi}\right)-2\sqrt{Q_{1}Q_{5}}\sqrt{r^
{2}+1}
\sin\theta \rho\left(\tilde{\phi}\right)
\cos\left(\phi-\tilde{\phi}\right)
\end{equation*}
In terms of the new variables the vacuum is given by
\begin{equation*}
\rho_{0}=\sqrt{Q_{1}Q_{5}},\, u_{0}=Q_{5}\phi
\end{equation*}

\section{Appendix B}
\subsection{Details of the evaluation of the symplectic form}
The non zero components of the inverse of the metric are given by
\begin{eqnarray*}
g^{tt}=\frac{-f_{1}f_{5}+A^{2}}{\sqrt{f_{1}f_{5}}},\,\,
g^{yy}=\frac{f_{1}f_{5}+B^{2}}{\sqrt{f_{1}f_{5}}},\,\ \\
g^{x^{i}x^{j}}=\delta^{ij}\frac{1}{\sqrt{f_{1}f_{5}}},\,\
g^{yx^{i}}=-\frac{B^{i}}{\sqrt{f_{1}f_{5}}},\,\,
g^{t^{x_{i}}}=\frac{A_{i}}{\sqrt{f_{1}f_{5}}}
\end{eqnarray*}
and the determinant is given by
\begin{equation*}
g=-f_{1}f_{5}.
\end{equation*}
It is useful to see the components of the matrix
\begin{equation*}
j^{mn}=\sqrt{-g}g^{mn}
\end{equation*}
which are given by
\begin{eqnarray*}
j^{tt}=-f_{1}f_{5}+A^{2},\,\,
j^{yy}=f_{1}f_{5}+B^{2}\\
j^{x^{i}x^{j}}=\delta^{ij},\,\ j^{yx^{i}}=-B^{i},\,\, j^{t
x_{i}}=A_{i}
\end{eqnarray*}
The nonzero Christoffel symbols that we will need are given by
\begin{eqnarray*}
\Gamma^{t}_{tt}&=&-\frac{A_{i}\partial_{i}\sqrt{w}}{2\sqrt{w}^{3}}\\
\Gamma^{t}_{yy}&=& \frac{A_{i}\partial_{i}\sqrt{w}}{2\sqrt{w}^{3}}\\
\Gamma^{t}_{x^{i}t}&=& \frac{1}{2\sqrt{w}^{3}}
\left[\sqrt{w}A_{j}\partial_{i}A_{j}-\sqrt{w}A_{j}\partial_{j}A_{i}+
A_{i}A_{j}\partial_{j}\sqrt{w}-
\sqrt{w}^{2}\partial_{i}\sqrt{w} \right]\\
\Gamma^{t}_{x^{i}y}&=&\frac{1}{2\sqrt{w}^{3}}
\left[\sqrt{w}A_{j}\partial_{i}B_{j}-\sqrt{w}A_{j}\partial_{j}B_{i}+
B_{i}A_{j}\partial_{j}\sqrt{w} \right]\\
\Gamma^{n}_{nx^{i}}&=& \frac{\partial_{i}\sqrt{w}}{\sqrt{w}}
\end{eqnarray*}
Where we have set $w=f_{1}f_{5}$. The gravitational contribution
from the bulk gives
\begin{eqnarray*}
J^{t}_{G}=&&
-\delta\left(\frac{A_{i}\partial_{i}\sqrt{w}}{\sqrt{w}^{3}}
\right)\wedge \delta\left(w
\right)-\delta\left(\frac{A_{i}\partial_{i}\sqrt{w}}{2\sqrt{w}^{3}}
\right)\wedge \delta A^{2}+
\delta\left(\frac{A_{i}\partial_{i}\sqrt{w}}{2\sqrt{w}^{3}}
\right)\wedge \delta B^{2}\\
&&-\delta\left[\frac{1}{\sqrt{w}^{3}}
\left[\sqrt{w}A_{j}\partial_{i}A_{j}-\sqrt{w}A_{j}\partial_{j}A_{i}+
A_{i}A_{j}\partial_{j}\sqrt{w}- \sqrt{w}^{2}\partial_{i}\sqrt{w}
\right] \right]\wedge \delta A_{i}\\
&&+\delta\left[\frac{1}{\sqrt{w}^{3}}
\left[\sqrt{w}A_{j}\partial_{i}B_{j}-\sqrt{w}A_{j}\partial_{j}B_{i}+
B_{i}A_{j}\partial_{j}\sqrt{w} \right] \right]\wedge \delta B_{i}\\
&&+ \delta\left(\frac{\partial_{i}\sqrt{w}}{\sqrt{w}} \right)\wedge
\delta A_{i}
\end{eqnarray*}
After a little simplification we have the form
\begin{eqnarray*}
J^{t}_{G}=&&
-\delta\left(\frac{A_{i}\partial_{i}\sqrt{w}}{\sqrt{w}^{3}}
\right)\wedge \delta\left(w\right)\\
&&-\delta\left[\frac{1}{\sqrt{w}^{3}}
\left[\sqrt{w}A_{j}\partial_{i}A_{j}-gA_{j}\partial_{j}A_{i}
\right] \right]\wedge \delta A_{i}\\
&&+\delta\left[\frac{1}{\sqrt{w}^{3}}
\left[\sqrt{w}A_{j}\partial_{i}B_{j}-\sqrt{w}A_{j}\partial_{j}B_{i}
\right] \right]\wedge \delta B_{i}\\
&&+ 2\delta\left(\frac{\partial_{i}\sqrt{w}}{\sqrt{w}} \right)\wedge
\delta A_{i}
\end{eqnarray*}
The final result for the case of the gauge field part reads
\begin{eqnarray*}
J_{H}^{t}=&&-\frac{1}{2}
\delta\left[2\frac{1}{f_{5}}\partial_{\left[i\right.}B_{\left.j\right]}
+\frac{3\partial_{\left[i\right.}\mathcal{C}_{\left.
ja\right]}}{f_{5}^{2}}A^{a}
\right]\wedge\delta\left[\mathcal{C}_{ij}+\frac{2}{f_{1}}B_{\left[i
\right.}A_{\left.j\right]}
\right]\\
&&-\delta\left[-f_{1}^{2}\partial_{i}\frac{1}{f_{1}}
+\frac{1}{f_{5}}\left[B^{b}\partial_{i}B_{b}-B^{b}\partial_{b}B_{i}-A^
{a}\partial_{i}A_{a}
+A^{a}\partial_{a}A_{i}\right]-
\frac{1}{f_{5}^{2}}3\partial_{\left[a\right.}
\mathcal{C}_{\left.bi\right]}A^{a}B^{b}\right] \wedge
\delta\left[\frac{A_{i}}{f_{1}} \right]
\end{eqnarray*}
For the contribution of the scalar we simply have
\begin{equation*}
J_{a}^{t}=-\frac{1}{2}\delta\left(A^{i}\partial_{i}\ln\frac{f_{1}}{f_
{5}}
\right)\wedge \delta\ln\frac{f_{1}}{f_{5}}.
\end{equation*}

Putting everything together we have that the bulk gives
\begin{eqnarray*}
J_{bulk}^{t}=\int d^{4}x &&
\left\{-\frac{1}{2}\delta\left(A^{i}\partial_{i}\ln\frac{f_{1}}{f_{5}}
\right)\wedge \delta\ln\frac{f_{1}}{f_{5}}+\delta\left[ \frac{1}{f_{5}
^{2}}3\partial_{\left[a\right.}
\mathcal{C}_{\left.bi\right]}A^{a}B^{b} \right]\wedge
\delta\left[\frac{A_{i}}{f_{1}} \right] \right.\\
&&-\frac{1}{2}
\delta\left[2\frac{1}{f_{5}}\partial_{\left[i\right.}B_{\left.j\right]}
+\frac{3\partial_{\left[i\right.}\mathcal{C}_{\left.
ja\right]}}{f_{5}^{2}}A^{a}
\right]\wedge\delta\left[\mathcal{C}_{ij}+\frac{2}{f_{1}}B_{\left[i
\right.}A_{\left.j\right]}
\right]\\
&&-\delta\left[-f_{1}^{2}\partial_{i}\frac{1}{f_{1}}
+\frac{1}{f_{5}}\left[B^{b}\partial_{i}B_{b}-A^{a}\partial_{i}A_{a}
+A^{a}\partial_{i}A_{a}\right]\right] \wedge
\delta\left[\frac{A_{i}}{f_{1}} \right]\\
&& -\delta\left(\frac{A_{i}\partial_{i}\sqrt{w}}{\sqrt{w}^{3}}
\right)\wedge \delta\left(w\right)+ 2\delta\left(\frac{\partial_{i}
\sqrt{w}}{\sqrt{w}} \right)\wedge
\delta A_{i}\\
&&-\delta\left[\frac{1}{\sqrt{w}^{3}}
\left[\sqrt{w}A_{j}\partial_{i}A_{j}-gA_{j}\partial_{j}A_{i}
\right] \right]\wedge \delta A_{i}\\
&&\left. +\delta\left[\frac{1}{\sqrt{w}^{3}}
\left[\sqrt{w}A_{j}\partial_{i}B_{j}-\sqrt{w}A_{j}\partial_{j}B_{i}
\right] \right]\wedge \delta B_{i}\right\}
\end{eqnarray*}
After many cancelations because of the non-commutative nature
\begin{equation*}
\delta A\wedge \delta B=-\delta B\wedge \delta A
\end{equation*}
and the linear equations
\begin{eqnarray*}
dA&=&-\ast_{4}dB\\
d\mathcal{C}&=&-\ast df_{5}
\end{eqnarray*}
we can bring the integrand to a total divergence form
\begin{equation*}
J_{bulk}^{t}=-\int d^{4}x \partial_{d} \left\{\frac{1}{2}
\epsilon_{ijad} \delta\left[\frac{A^{a}}{f_{5}}\right]\wedge \delta
\left(C_{ij}\right)\right\}.
\end{equation*}

\subsection{The total derivative terms}
It is interesting to see how a one dimensional integral is obtained from
the above surface integrals. At this point we would like to
introduce a surface of topology $S^{1}\times S^{2}$ of radius
surrounding the ring. Using Gauss' law we are able to reduce the
calculation of the total derivative to a surface integral on the
above mentioned surface. The embedding in the four dimensional space
spanned by $x^{i},i=1\ldots 4$ is
\begin{eqnarray*}
&&\vec{x}\left(u,\theta,\phi\right)=x_{\perp}\left(
                    \begin{array}{cccc}
                     \sin\theta
                     \frac{\dot{F_{2}}}{\left|\dot{F}\right|}+F_{1},
             & -\sin\theta \frac{\dot{F}_{1}}{\left|\dot{F}\right|}+F_
{2},
             & \cos\theta \cos\phi, & \cos\theta \sin\phi \\
                    \end{array}
                  \right)\\
&&u\, \epsilon \left[0, 2\pi Q_{5}\right] ,\,\theta\, \epsilon
\left[0, \pi\right],\, \phi\,\epsilon \left[0, 2\pi\right)
\end{eqnarray*}
From the above we see that the measure of integration on the tube is
given by
\begin{equation*}
dS=x_{\perp}^{2}\left|\dot{F}\right| \sin\theta du\,d\theta\,d\phi
\end{equation*}
and the normal unit vector is
\begin{equation*}
\vec{n}\left(u,\theta,\phi\right)=\left(
                    \begin{array}{cccc}
                     \sin\theta
                     \frac{\dot{F_{2}}}{\left|\dot{F}\right|},
             & -\sin\theta \frac{\dot{F}_{1}}{\left|\dot{F}\right|},
             & \cos\theta \cos\phi, & \cos\theta \sin\phi \\
                    \end{array}
                  \right).
\end{equation*}
At the end of the calculation we want to consider the limit
$x_{\perp}\rightarrow 0^{+}$ which leads us to keeping the terms of
order $O\left(x_{\perp}^{-2} \right)$. We can indeed show that this
is the leading behavior of the integrand.
\par
From the expressions that were previously given we have the
variations
\begin{eqnarray*}
\delta f_{5}&=&\frac{2Q_{5}}{L}\int_{0}^{L}\frac{du}{\left|x-F
\right|^{4}}\left(x_{i}-F_{i}\right)\delta F_{i}\\
\delta f_{1}&=&\frac{Q_{5}}{L}\int_{0}^{L}\frac{\delta\dot{\left|F
\right|}^{2}du}{\left|x-F \right|^{2}} +
\frac{2Q_{5}}{L}\int_{0}^{L}\frac{\dot{\left|F
\right|}^{2}du}{\left|x-F
\right|^{4}}\left(x_{i}-F_{i}\right)\delta F_{i} \\
\delta A_{j}&=&-\frac{Q_{5}}{L} \int_{0}^{L}du\frac{\delta
F_{j,u}}{\left|x-F \right|^{2}}-\frac{2Q_{5}}{L}
\int_{0}^{L}du\frac{F_{j,u}}{\left|x-F
\right|^{4}}\left(x_{i}-F_{i}\right)\delta F_{i}
\end{eqnarray*}
We would like to approximate the above expressions in the limit
where $\mathbf{x}$ approaches some point on the curve
$F\left(u\right)$
\begin{eqnarray*}
\delta f_{5}&\approx&\frac{1}{2\dot{\left|F
\right|}{x^{\perp}}^{2}}n_{i}\delta F_{i} + O\left(x_{\perp}^{-1}
\right)\\
\delta f_{1}&\approx&\frac{\dot{\left|F
\right|}}{2{x^{\perp}}^{2}}n_{i}\delta F_{i} + O\left(x_{\perp}^{-1}
\right) \\
\delta A_{j}&\approx&- \frac{\dot{F}_{j}}{2\dot{\left|F
\right|}{x^{\perp}}^{2}}n_{i}\delta F_{i}- \frac{\delta
\dot{F}_{j}}{2\dot{\left|F \right|}{x^{\perp}}}+
O\left(x_{\perp}^{0}\right).
\end{eqnarray*}
We can also use the asymptotic expansions
\begin{eqnarray*}
f_{1}&\approx& \frac{\dot{\left|F \right|}}{2{x^{\perp}}}\\
f_{5}&\approx& \frac{1}{2\dot{\left|F \right|}{x^{\perp}}}\\
A_{i}&\approx&-\frac{\dot{F}_{i}}{2\dot{\left|F \right|}{x^{\perp}}}
\end{eqnarray*}
which give the useful identity
\begin{equation*}
n_{i}g^{tx^{i}}= - \frac{n_{i}\dot{F}_{i}}{2\dot{\left|F \right|}}+
O\left(x_{\perp}\right)=O\left(x_{\perp}\right)
\end{equation*}
The above expansion enables us to determine the asymptotic behavior
of the covariant derivative also. The contribution of the gauge
transformation term is given by
\begin{eqnarray*}
&&\int d^{4}x \partial_{d} \left\{\sqrt{-g}g^{dm}g_{rq}
\nabla_{m}\delta g^{r t} \wedge\xi^{q}-
\sqrt{-g}\nabla_{m}\xi^{d}\wedge\delta g^{mt}\right\}\\
&&=2\pi \int_{0}^{L} du \frac{1}{\left|
\dot{F}\right|^{2}}\left(\dot{F}_{2} \delta \dot{F}_{1} -\dot{F}_{1}
\delta \dot{F}_{2} \right)\wedge \left(\dot{F}_{2} \delta F_{1} -
\dot{F}_{1} \delta F_{2}  \right)
\end{eqnarray*}
The calculation of the term coming from the bulk is a little more
involved. The equation that needs to be solved for $C_{ij}$ is a
monopole type equation. As usually we will need to solve the equation
on patches. In order to make the
calculation easier we will define the fourth direction to be the
direction defined by the tangent on the ring at some point
$F\left(u\right)$. The gauge that we fix is
\begin{equation*}
C_{4i}=0.
\end{equation*}
The equations now take the form
\begin{eqnarray*}
\partial_{4}C_{\alpha \beta}&=& \epsilon_{\alpha \beta
\gamma}\partial_{\gamma}f_{5}\\
\frac{1}{2}\epsilon_{\alpha \beta \gamma}\partial_{\alpha}C_{\beta
\gamma}&=&\partial_{4}f_{5}
\end{eqnarray*}
Where $\alpha,\beta,\gamma=1,2,3$. For the variation we have
\begin{eqnarray*}
\partial_{4}\delta C_{\alpha \beta}&=& \epsilon_{\alpha \beta
\gamma}\partial_{\gamma}\delta f_{5}\\
\frac{1}{2}\epsilon_{\alpha \beta \gamma}\partial_{\alpha} \delta
C_{\beta \gamma}&=&\partial_{4}\delta f_{5}.
\end{eqnarray*}
Solving the equations close to the ring yields
\begin{equation*}
\delta C_{\alpha \beta}= \epsilon_{\alpha \beta \gamma}
\frac{n_{\gamma}\delta F}{2x_{\perp}}
\end{equation*}
where
\begin{equation*}
\delta F=\frac{\dot{F}_{i}\delta{F}_{i}}{\left|\dot{F} \right|}
\end{equation*}
is the variation parallel to the ring. The integral that we need to
perform now reads
\begin{eqnarray*}
\int d^{4}x \partial_{d}\left[ \frac{1}{2} \epsilon_{ijad}
\delta\left[\frac{A^{a}}{f_{5}}\right]\wedge \delta
\left(C_{ij}\right)\right]= -2\pi \int_{0}^{L} du \frac{1}{\left|
\dot{F}\right|^{2}}\left(\dot{F}_{1} \delta \dot{F}_{1} +
\dot{F}_{2} \delta \dot{F}_{2} \right)\wedge \left(\dot{F}_{1}
\delta F_{1} + \dot{F}_{2} \delta F_{2}  \right)
\end{eqnarray*}

\end{document}